\documentclass[english,number,preprint,review]{elsarticle}
\usepackage[T1]{fontenc}
\usepackage[latin9]{inputenc}
\usepackage{geometry}
\usepackage{mathtools}
\usepackage{amsmath}
\usepackage{amssymb}
\usepackage{graphicx}
\usepackage{esint}

\makeatletter

\providecommand{\tabularnewline}{\\}

\usepackage{dsfont,bbold,mathptmx,times,romanbar,datetime,url,siunitx}

\settimeformat{ampmtime}

\date{}

\newcommand{\prob}[1]{\mathbb{P}\left(#1\right)}

\newcommand{\ex}[1]{\mathbb{E}\left[#1\right]}

\newcommand{\exnobr}[1]{\mathbb{E}#1}

\newcommand{\Sustain}{\emph{Sustain}}

\newcommand{\one}{\mathbb{1}}

\newcommand{\vertTenPt}{\rule{0pt}{10pt}}

\def\D{\mathrm{d}}

\makeatother

\usepackage{babel}
\begin{document}

\begin{frontmatter}{}

\title{Model Predictive HVAC Control with Online Occupancy Model}

\author{Justin R. Dobbs\corref{cor}}

\ead{jrd288@cornell.edu}

\author{Brandon M. Hencey\corref{cor2}}

\ead{bmh78@cornell.edu}

\address{Department of Mechanical and Aerospace Engineering, Upson Hall, Cornell
University, Ithaca, NY 14853, USA}

\cortext[cor]{Corresponding author. Tel +1 (607) 269--5352.}
\begin{abstract}
This paper presents an occupancy-predicting control algorithm for
heating, ventilation, and air conditioning (HVAC) systems in buildings.
It incorporates the building's thermal properties, local weather predictions,
and a self-tuning stochastic occupancy model to reduce energy consumption
while maintaining occupant comfort. Contrasting with existing approaches,
the occupancy model requires no manual training and adapts to changes
in occupancy patterns during operation. A prediction-weighted cost
function provides conditioning of thermal zones before occupancy begins
and reduces system output before occupancy ends. Simulation results
with real-world occupancy data demonstrate the algorithm's effectiveness.
\end{abstract}

\end{frontmatter}{}

\section{Introduction}

\thispagestyle{empty}The long-term increase in energy prices has
driven greater interest in demand-based HVAC control. Fixed temperature
setpoint schedules and occupancy-triggered operation are commonly
used to trim energy consumption, but these approaches have significant
drawbacks. First, fixed schedules become outdated; when occupancy
patterns change, early or late occupants are left uncomfortable, or
the space is conditioned prematurely or for too long. Second, thermal
lag limits response speed and thus precludes aggressive temperature
set-back. Addressing both schedule inaccuracy and thermal lag requires
a stochastic occupancy model and a control scheme that can use it
effectively.

Considerable research effort has been directed toward occupancy detection
and modeling. Work on detection has focused on boosting accuracy through
sensor fusion using probabilistic, neural, or utility networks \citep{lam2009occupancy,dodier2006building,meyn2009sensor,modelingCountData}.
Agent-based models have been used to predict movement within buildings
\citep{liao2012agent,erickson2009energy}, as have Markov chains \citep{erickson2010occupancy,page2008generalised,dong2011integrated}.
Erickson and Dong, for example, considered rooms to be Markov states
and movements among them to be transitions in order to predict persons'
behavior, while Dong and Lam \citep{dong2014real} used a semi-Markov
model to merge multiple sensor streams into an occupant count estimate.
The simpler Page model considered boolean occupancy (occupied or vacant)
under a time-heterogeneous Markov chain to generate realistic simulation
input data, rather than for on-line forecasting \citep{page2008generalised}.

With the exception of the Page model, the above efforts have found
use in heuristic \citep{erickson2011observe,selfprogthermostat,goyal2013occupancy}
or model predictive control (MPC) schemes \citep{dong2014real,Oldewurtel201215,goyal2013occupancy},
but they face barriers to widespread usage. Most notably, where authors
have used MPC, they have also used manually-generated thermal models
\citep{dong2014real,Oldewurtel201215,goyal2013occupancy} even though
model creation is tedious and time-consuming and therefore expensive.
Eager to demonstrate excellent performance, researchers have favored
systems with complex topologies and numerous adjustments that yield
``one-off'' engineering efforts without a clear path to large-scale
adoption. The system outlined in \citep{dong2014real}, for instance,
uses $\textrm{CO}_{2}$, sound, and light sensors that require carefully
set detection thresholds for each room, plus an on-board weather forecasting
algorithm in lieu of forecasts already available. We aim, instead,
to make occupancy-predicting control accessible to a broader audience
by presenting a simple but effective algorithm with a straightforward
implementation. For example, we use an automated BIM translation facility
outlined in a previous paper \citep{Greenberg201344}, and the core
algorithm is industry-standard {MPC} with occupancy weighting in
the cost function. Each of the very few adjustments serves a clearly-defined
purpose, and we have outlined each component's operation with the
practitioner in mind. 

Second, recent research has paid little attention to the commissioning
and maintenance of occupancy prediction algorithms; model training,
if mentioned at all, has been a secondary consideration assumed to
by done one time by someone skilled in the art \citep{page2008generalised,dong2014real,erickson2011observe}.
Although most training algorithms could be extended to work on-line,
ongoing maintenance remains a source of long-term cost neglected by
the literature. An occupancy model invariably becomes out-of-date
unless it is periodically retrained or can incrementally refine itself
with new observations. Our work uses on-line Bayesian inference for
stable performance without ongoing manual effort.

\begin{figure*}
\begin{centering}
\includegraphics[scale=0.25]{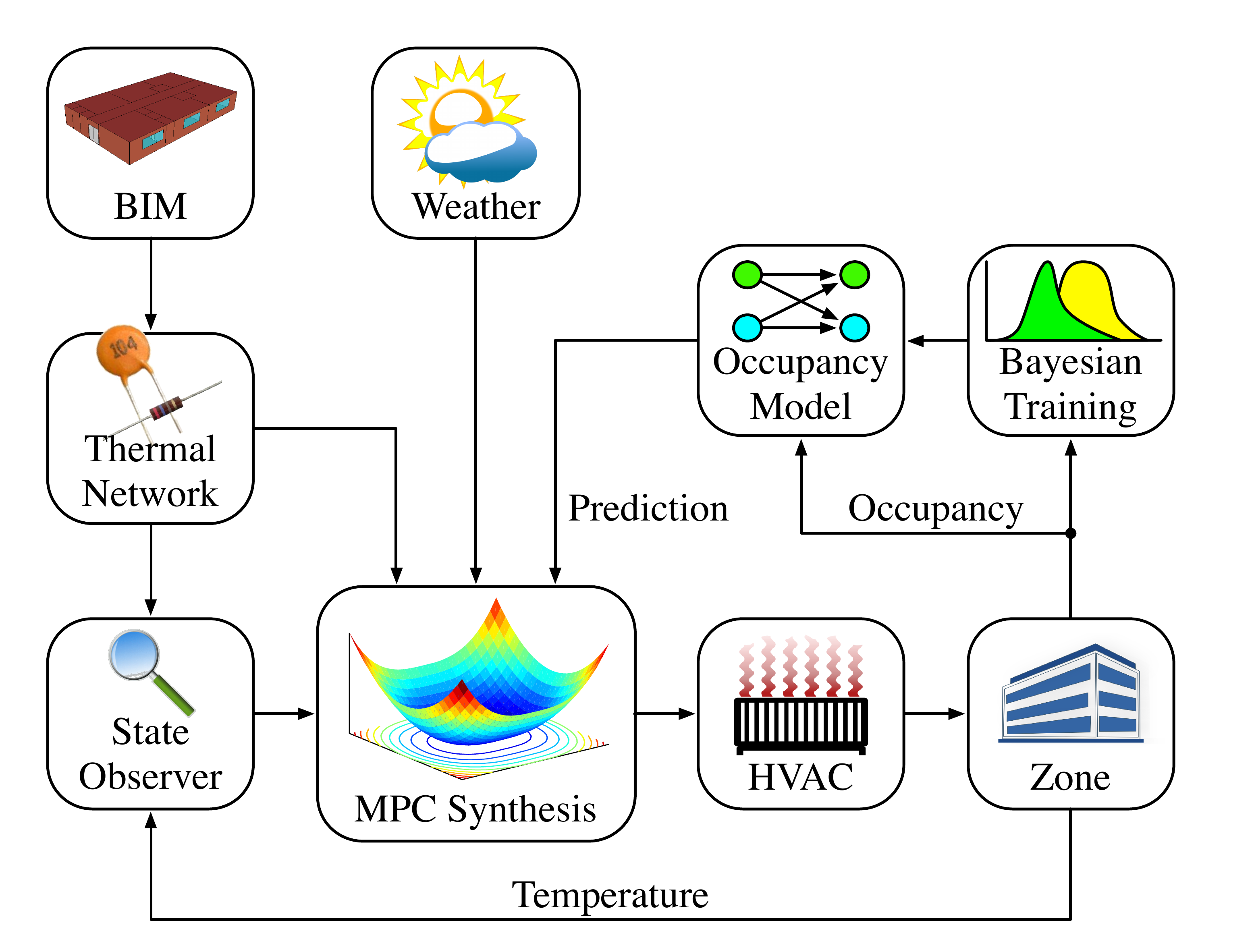}
\par\end{centering}

\protect\caption{\label{fig:ProposedArch}Proposed system architecture. For this study,
the building model has been translated automatically from CAD data
into a linear, time-invariant network that encompasses the dominant
thermal processes. (Model translation may also be performed manually.)}
\end{figure*}

The paper progresses as follows. First, we outline the problem formulation.
Second, we describe the stochastic occupancy model and its on-line
training algorithm. Third, we discuss its integration with model predictive
control. Finally, we present simulation results using real-world occupancy
data and compare our method's performance to a correctly set scheduled
controller and to an occupancy-triggered controller. Throughout the
discussion, the control scenario is kept deliberately simple to emphasize
the contribution of occupancy learning and its use with MPC.%
\footnote{See \citep{goyal2013occupancy} for a comparison of MPC and heuristic
control for a more complex HVAC system.%
}

\section{\label{sec:Problem-Statement}Problem Statement}

\begin{figure}
\begin{centering}
\includegraphics[scale=0.25]{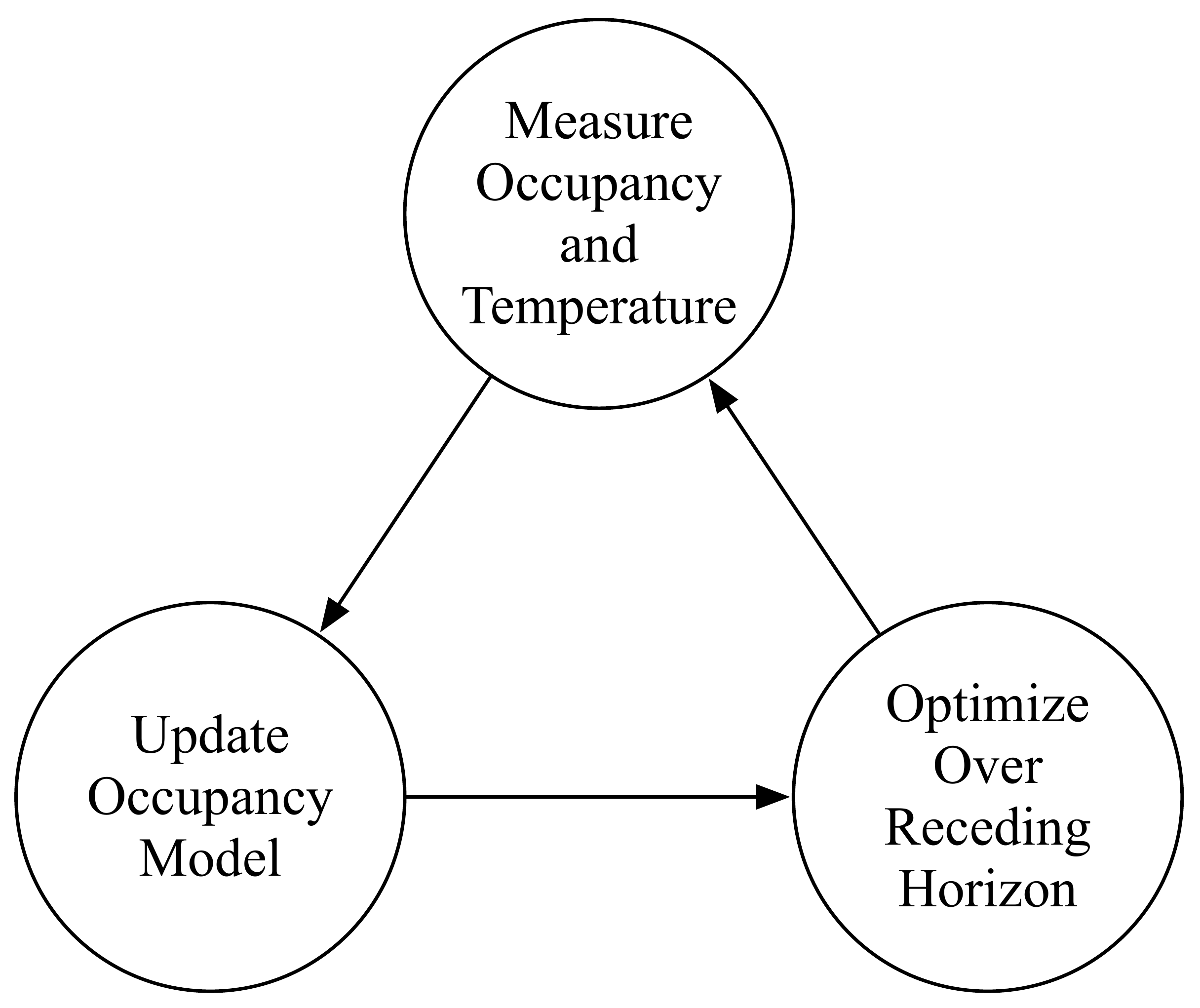}
\par\end{centering}

\protect\caption{\label{fig:Process-flow-during}Process flow during operation.}
\end{figure}

We wish to minimize the total energy usage of a building heating (or
cooling) system while maintaining occupant comfort. Versus conventional
occupancy-triggered or scheduled control, we aim to 
\begin{itemize}
\item boost comfort by conditioning the space before occupants arrive,
\item limit energy consumption by not running the system too early, and
\item exploit stored thermal energy by reducing power before occupants leave.
\end{itemize}
Our approach is based on MPC but uses a cost function weighted by
occupancy predictions from a self-training stochastic model (Figure~\ref{fig:ProposedArch}).
At each step, the system measures how much of the previous hour the
space was occupied, and the expected occupancy is used to find the
best sequence of $N$ future heat inputs to the thermal zone that
minimizes the expected cost. The optimization is
\begin{equation}
\begin{aligned}\min_{u_{k}\cdots u_{k+N-1}}\quad & \sum_{j=0}^{N-1}\ex{g(x_{k+j},u_{k+j},\tau,\Gamma_{k+j})}\\
\textrm{subject to\quad} & \begin{aligned}x_{i+1} & =Ax_{i}+B_{u}u_{i}+B_{w}\ex{w_{i}}\quad\forall i\in\mathbb{Z}^{+}\\
0 & \leq u\leq u_{\max}
\end{aligned}
\end{aligned}
\label{eq:firstMpcDef}
\end{equation}
 where
\begin{itemize}
\item $k\in\mathbb{Z}^{+}$ is the current time step, and $j\in[0,N-1]$
is the optimization index over the horizon;
\item $A\in\mathbb{R}^{n\times n}$ describes the building's thermal dynamics;
\item $x\in\mathbb{R}^{n\times1}$ contains the building's thermal state;
\item $u_{k\ldots k+N-1}$ contains the controller output, constrained within
the system's capacity $u_{\max}$;
\item $B_{u}$ is a vector that connects the heat input $u$ to the zone
air volume;
\item $w_{k}$ is the current weather observation, and $w_{k+1\ldots k+N-1}$
contains an up-to-date weather prediction;
\item $B_{w}$ is a vector that connects the weather conditions to the building
envelope;
\item $\tau$ is the temperature setpoint, which is constant for this study
(but can be varied in practice);
\item $\Gamma_{k}$ is the latest occupancy measurement, and $\Gamma_{k+1\ldots k+N-1}$
are the predicted occupancies; and
\item $g(x,u,\tau,\Gamma)$ is a cost function that penalizes total energy
consumption and penalizes discomfort based on the occupancy $\Gamma$.
\end{itemize}
The expectation operator $\ex{g}$ in Equation~\ref{eq:firstMpcDef}
reflects that future values of $g$ require predictions of occupancy
and of the weather. The optimization yields an optimal sequence of
$N$ power commands to the HVAC system, where positive values are
heat and negative are cooling; the first command $u_{k}$ is applied,
and the rest are discarded. The previous and current occupancy observations
are then used to train the occupancy model, and the entire process
repeats the next time step (Figure~\ref{fig:Process-flow-during}).

Two assumptions are made in this presentation. First, we treat the
weather forecast as accurate so that we can later omit the expectation
operator from $w$. Second, we use a very simple cost function with
constant efficiency and a single linear actuator. These assumptions
improve clarity but are not required in practice. Where available,
weather uncertainty data can be rolled into the cost function in order
to improve robustness \citep{Oldewurtel201215}. Multiple actuators
(e.g.\ radiant and forced air with vastly different response times)
or nonlinear actuation (e.g.\ variable air volume damper position)
can be pulled into the dynamical model and the cost function without
undermining the basic approach \citep{haves2010model,Oldewurtel201215,goyal2013occupancy}.
Finally, the energy penalty gain can be varied over time to reflect,
for example, changing system efficiency or electricity cost.

\section{Building Thermal Model}

Thermal model accuracy influences controller performance, so we need
a thermal model that closely approximates the dominant dynamics. Here
we outline how the state-space building model is generated, and we
validate it against EnergyPlus simulation results.

\begin{figure}
\begin{centering}
\includegraphics[width=90mm]{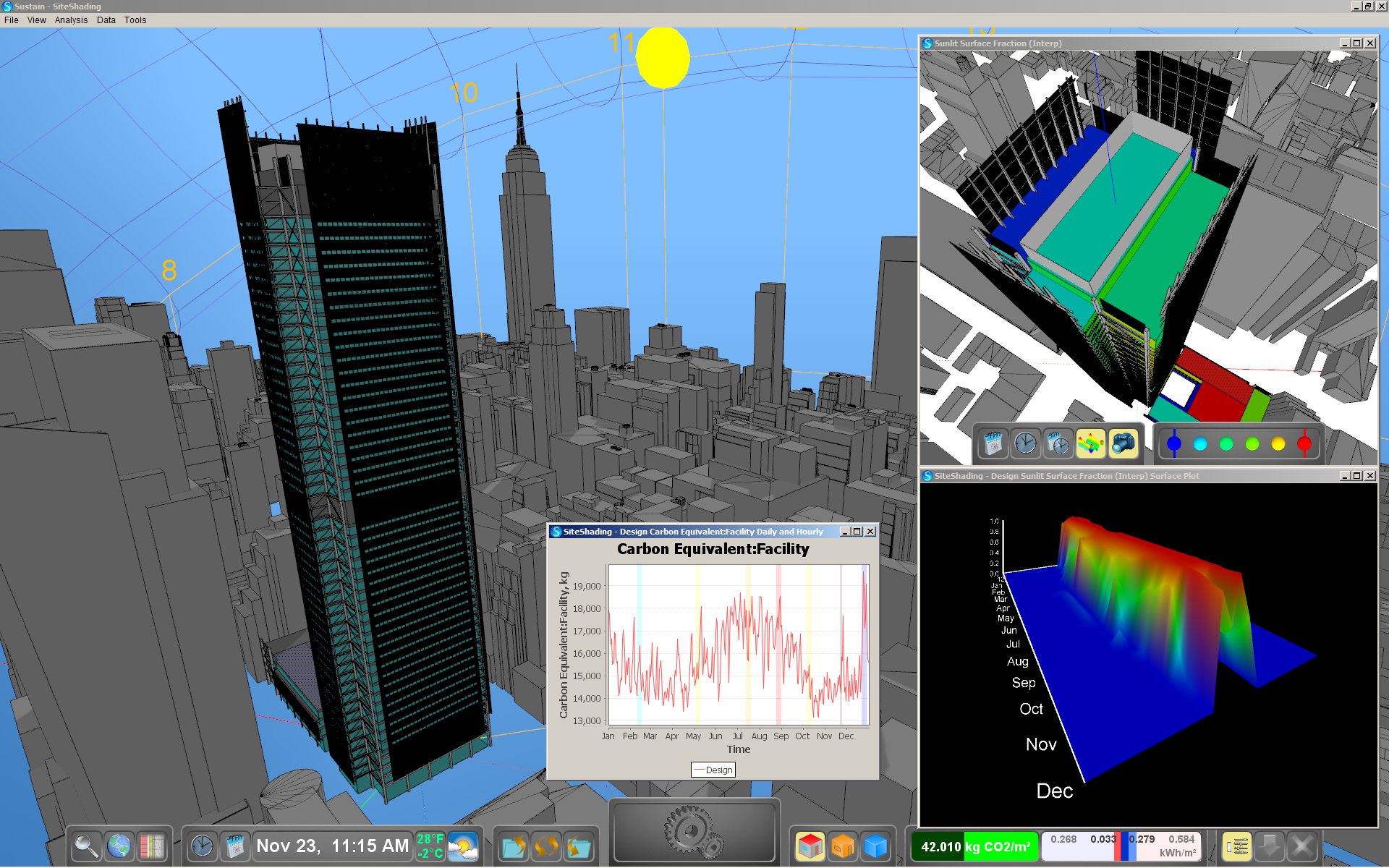}
\par\end{centering}

\protect\caption{\label{fig:Sustain}The \Sustain{} modeling and simulation environment.
(New York Times building shown.)}
\end{figure}

Thermal model creation has historically been a manual process contributing
substantially to MPC implementation cost. Research efforts such as
the \Sustain{} platform (Figure~\ref{fig:Sustain}) \citep{Greenberg201344,dobbsautomatic}
and the Building Resistance-Capacitance Modeling Toolbox \citep{sturzenegger2014brcm}
have arisen to streamline the creation of dynamical equations suitable
for MPC\@. Here we have used a module in \Sustain{} to generate
a resistor-capacitor network directly from a CAD model. The thermal
model states are the building's internal temperatures, including zone
air plus wall layers and roofing materials that are not normally measured;
a state observer can easily estimate these values during operation.%
\footnote{The observability assumption is valid because of the RC network's
construction; the driving sources (exterior and interior conditions)
are themselves measurable and there are no hinges in the network.
See \citep{dobbsautomatic} for details on the RC network construction
and \citep{luluControllabilityAndObservability,dobbsautomatic} Theorem~1
for a proof of observability.%
} Although not used here, ways to automatically tune the RC network
parameters on-line and even estimate disturbances such as solar load
have recently been introduced \citep{radecki_online_2013}.

The model used for this study has 41 states: one for zone air and
the rest for building structure%
\footnote{The model can include multiple control zones if needed. In practice,
one may reduce the model size using balanced truncation, which reconfigures
the state space. We have chosen to retain the full-order model to
maximize accuracy and preserve physical intuition. See \citep{dobbscdc2012}
for a survey of methods that reduce state space size while preserving
structure.%
}. It assumes well-mixed air and uses time-invariant convection coefficients.
Fixed coefficients imply that the thermal gradients are always in
the same direction, whereas EnergyPlus switches coefficients depending
on whether the gradient enhances convection \citep{EnergyPlusReference2011}.
In practice, for improved accuracy, the RC network can be adjusted
at each step, or a nonlinear model may be used. We have included limited
support for radiant transfer using coefficients from EnergyPlus' \texttt{Simple}
and \texttt{SimpleCombined} convection algorithms \citep{EnergyPlusReference2011}.
 The model accepts the following inputs:
\begin{itemize}
\item the outside dry-bulb temperature,
\item the ground temperature, and%
\footnote{Daily ground temperature is available for free through on-line sources
such as the U.S. Surface Climate Observing Reference Network \citep{noaaClimateData}.%
}
\item heat injected by the control system to the space (positive or negative).
\end{itemize}
The state equation of the building is
\begin{equation}
x_{k+1}=A_{b}x_{k}+B_{w}w_{k}+B_{u}u_{k},\label{eq:buildingStateEquation}
\end{equation}
where $k$ is the time step (in hours) and $x_{k}$ is the complete
temperature state vector containing the zone temperature $x_{k}^{\textrm{zone}}$.
The vector $w_{k}$ contains the weather forecast, and $u_{k}$ is
the heat injected into the room by the HVAC system. The sign of $u_{k}$
and its constraints can be made negative, or the sign of the vector
$B_{u}$ can be reversed, for cooling.

\begin{figure}
\begin{centering}
\includegraphics{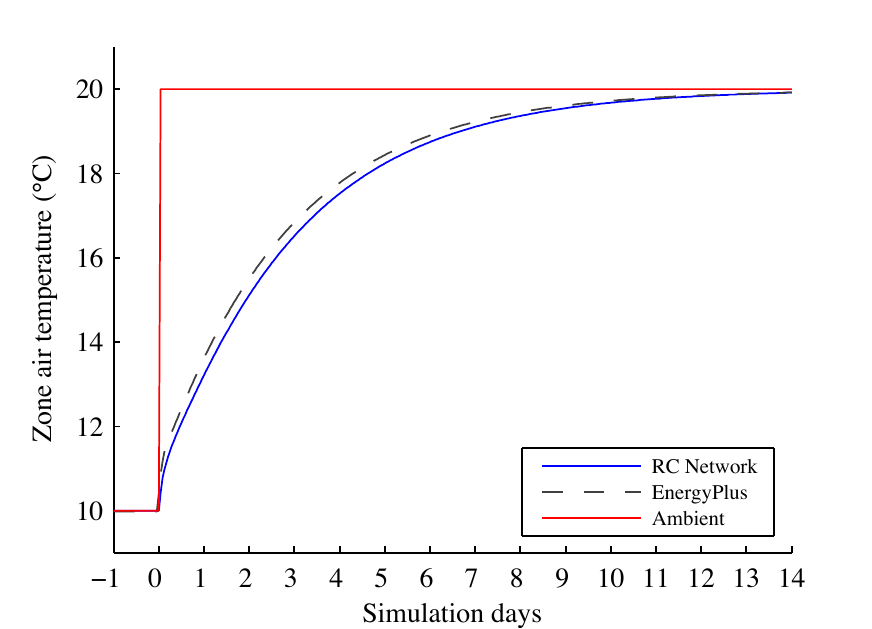} 
\par\end{centering}

\protect\caption{\label{fig:rcVersusEp}RC network (solid) and EnergyPlus (dashed)
simulation results for a step change in ambient temperature.}

\end{figure}

Let us now validate the model by comparing the zone temperature time
response of the RC network to EnergyPlus results under simplified
conditions. The goal is not to exactly match EnergyPlus, but rather
to show that the dominant response is plausibly close. To do this,
we have simulated the building using first the RC network and then
EnergyPlus under the following set of conditions:
\begin{itemize}
\item a step change in air, ground, and sky infrared temperatures from 10$^{\circ}\textrm{C}$
to 20$^{\circ}\textrm{C}$,
\item no wind or humidity, and
\item EnergyPlus heat transfer algorithms: \texttt{Simple} convection for
interior, \texttt{SimpleCombined} for exterior, and \texttt{CTF} (conduction
transfer function) for walls.
\end{itemize}
The RC network implementation lacks support for sky infrared transfer
through windows; by matching the sky radiant temperature to the outside
air temperature, we have removed this source of discrepancy from the
simulation.  Under the simplified conditions, very similar response
times (Figure~\ref{fig:rcVersusEp}) suggest that the RC model is
adequate for demonstration.

\subsection*{}

\section{\label{sec:Stochastic-occupancy-model}Stochastic Occupancy Model}

The heart of our method is its on-line trained Markov occupancy model
that quickly adapts and enables the MPC to predict occupancy. The
input is a stream of asynchronous pulses from pyroelectric infrared
(PIR) or similar sensors that indicate whether at least one person
is in the space. We have chosen the Mitsubishi Electric Research Lab
(MERL) motion detector data set \citep{merldata}, which consists
of a series of one-second pulses from various motion sensors located
throughout hallways and conference rooms in MERL.%
\footnote{Although we have used the MERL occupancy data, the thermal model is
not one of the MERL building.%
} The meetings in the Belady conference room show a good balance between
repetition and variety to showcase the benefits of on-line learning.

\subsection{Markov Chain Formulation\label{sub:Markov-chain-formulation}}

\begin{figure*}
\centering{}\includegraphics[scale=0.25]{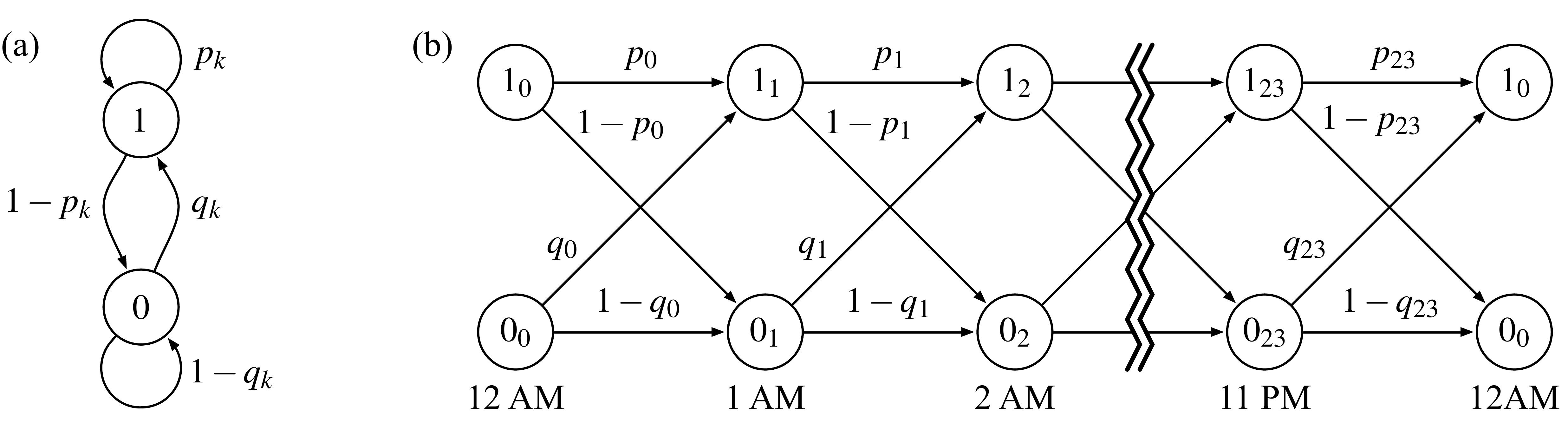}\protect\caption{\label{fig:markovChain}Occupancy model as a time-varying Markov chain
(a) and unrolled into a periodic structure (b).}
\end{figure*}
The occupancy model is as a periodic Markov chain updated at every
observation. The occupancy at time $k$ is either $\gamma_{k}=$1
(occupied) or $\gamma_{k}=0$ (vacant). The current occupancy state
and the time of day determine the probability of future occupancy.
We wish to estimate the probabilities 
\begin{equation}
\begin{aligned}p_{k} & =\prob{\gamma_{k+1}=1\mid\gamma_{k}=1},\\
q_{k} & =\prob{\gamma_{k+1}=1\mid\gamma_{k}=0}.
\end{aligned}
\end{equation}
The transition probabilities of this two-state time-varying Markov
chain (Figure~\ref{fig:markovChain}a) are periodic; we have chosen
a period $M=24$ hours, so $p_{24}\equiv p_{0}$ and $q_{24}\equiv q_{0}$.
To better visualize the periodicity, we unroll the Markov chain into
$2M$ states (Figure~\ref{fig:markovChain}b), where each hour has
a $1_{k}$ and $0_{k}$ state. Although $k$ in general grows without
bound, its range  is limited to $0\leq k\leq M-1$ when dealing with
the Markov chain. The choice of $M$ affects how learned patterns
relate to subsequent predictions; if space usage patterns vary significantly
across the weekdays, one might want to prevent occupancy observations
on Monday from influencing control actions on Tuesday, in which case
a one-week chain would be more appropriate. For this study, the one-day
Markov chain is trained using Monday through Friday occupancy data
from the MERL data set, ignoring weekends. In practice, one could
switch to a different Markov chain or use occupancy-triggered control
over weekends.

\subsection{Training}

\begin{figure}
\begin{centering}
\includegraphics[scale=0.25]{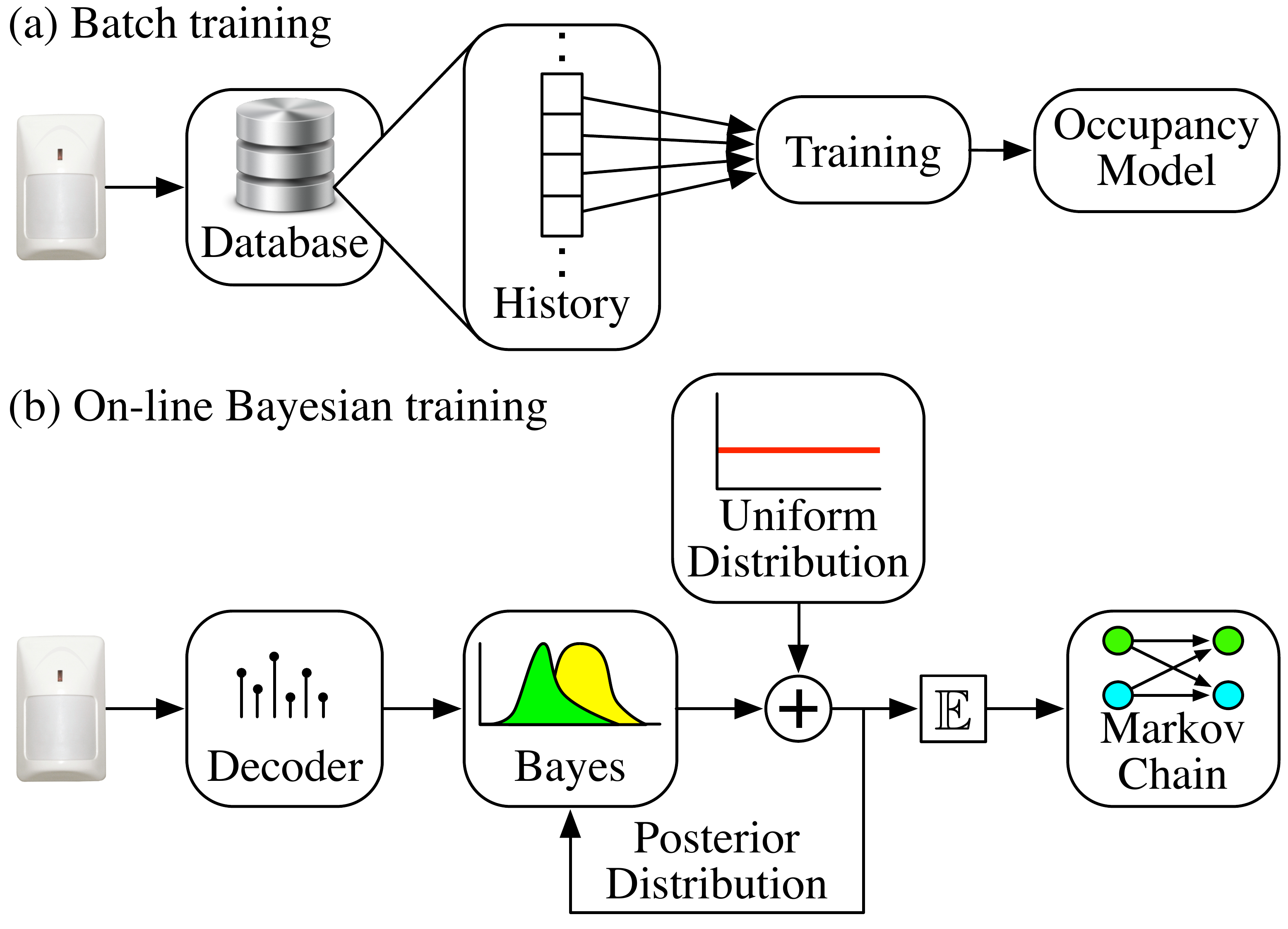}
\par\end{centering}

\protect\caption{\label{fig:trainingSchematic}Conventional batch training (a) versus
the proposed on-line incremental Bayesian training algorithm with
forgetting (b).}
\end{figure}

In contrast to batch training, which uses a fixed-size history (Figure~\ref{fig:trainingSchematic}a)
, on-line incremental training proceeds without user intervention.
It uses observations to update density functions for each of the transition
probabilities; the expected values of these density functions in turn
populate the Markov chain's transition matrix (Figure~\ref{fig:trainingSchematic}b).

Boolean occupancy lacks granularity that could otherwise make predictions
more accurate. For example, occupancy for the entirety of the previous
hour implies different future occupancy compared to just a few minutes.
The question we wish to answer is: \emph{Given the space was occupied
for a certain fraction of the previous hour, for what portion of subsequent
hours do we expect occupancy?} We approach the problem in three steps.
First, we explain the simplest case where boolean occupancy is directly
observed. Second, we augment the boolean training with forgetting
capability. Finally, we refine the approach to use fractional occupancy
in order to make predictions more precise.

\subsubsection{Boolean observed occupancy}

Each state of the unrolled Markov chain (Figure~\ref{fig:markovChain}b)
has two outgoing transition paths, analogous to a coin toss where
the coin's bias is unknown. The well-known probability function of
a biased coin is
\begin{equation}
\psi(\theta,N,N_{H})=\binom{N}{N_{H}}\theta^{N_{H}}(1-\theta)^{N-N_{H}},
\end{equation}
where $\binom{N}{N_{H}}$ is the number of ways to permute $N_{H}$
heads in a sequence of $N$ tosses, and $\theta$ is the heads bias
(with 0.5 being a fair coin). This function can be parameterized on
$\theta$, $N$, or $N_{H}$ depending on the purpose. With the bias
$\theta=\theta_{0}$ known and the number of tosses $N=N_{0}$ fixed,
the probability of obtaining $N_{H}$ heads, $\psi(\theta=\theta_{0},N=N_{0},N_{H})$,
is a discrete binomial distribution over $N_{H}$. When $N$ and $N_{H}$
are fixed, $\psi(\theta,N=N_{0},N_{H}=N_{H_{0}})$ is the probability
density over the bias $\theta$, with $\int_{0}^{1}\psi(\theta,N=N_{0},N_{H}=N_{H_{0}})\,\D\theta=1$.%
\footnote{The function $f(\theta)$ is a continuous beta distribution. Once
linear forgetting is added, these distributions become prohibitive
to maintain analytically because sums of beta distributions are not
themselves beta distributions, but rather are complicated piecewise
functions \citep{gupta2004handbook}. Therefore it is more practical
to maintain numerical approximations.%
}

Instead of computing $\psi$ using $N$ and $N_{H}$ all at once,
we can obtain it iteratively using Bayes' rule. Suppose we have a
sequence of outcomes $x_{j}\in\{1,0\}$ where $1$ means heads. The
distribution, now parameterized only on $\theta$, is defined recursively
as
\begin{equation}
\begin{aligned}\psi_{j}(\theta\mid x_{1\ldots j}) & \sim\psi_{j-1}(\theta\mid x_{1\ldots j-1})\Phi(\theta,x_{j})\\
 & =\frac{\psi_{j-1}(\theta\mid x_{1\ldots j-1})\Phi(\theta,x_{j})}{\int_{0}^{1}\psi_{j-1}(\theta\mid x_{1\ldots j-1})\Phi(\theta,x_{j})\D\theta},
\end{aligned}
\end{equation}
 where 
\begin{equation}
\Phi(\theta,x)=\begin{cases}
\theta & x=1\\
1-\theta & x=0,
\end{cases}
\end{equation}
and $\psi_{0}(\theta)=1$ is a uniform distribution reflecting no
prior knowledge of the bias. The $\sim$ notation means dividing by
a constant so that $\int_{0}^{1}\psi_{j}(\theta)\,\D\theta=1$ holds.
Our best guess of the bias is $\ex{\psi_{j}(\theta)}=\int_{0}^{1}\theta\psi_{j}(\theta)\D\theta$.

Now let us apply this analogy to occupancy prediction. Coin toss outcomes
are independent, but occupancy transition probabilities depend on
the current state. An any given time there are two possible states,
so we need to maintain two distributions per time step. Let $\gamma_{k}\in\{0,1\}$
be the occupancy. The transition probabilities of interest are
\begin{equation}
\begin{aligned}p_{k} & =\prob{\gamma_{k+1}=1\mid\gamma_{k}=1}=\ex{f_{k}(p_{k})}\\
q_{k} & =\prob{\gamma_{k+1}=1\mid\gamma_{k}=0}=\ex{g_{k}(q_{k})},
\end{aligned}
\end{equation}
where the density functions $f_{k}(p_{k})$ and $g_{k}(q_{k})$ are
the latest iterations of $f_{k,j}(p_{k})$ and $g_{k,j}(q_{k})$,
updated each training instance $j$ using
\begin{equation}
\begin{aligned}f_{k,j}(p_{k}\mid\gamma_{1\ldots k+1},\gamma_{k}=1) & \sim f_{k,j-1}(p_{k}\mid\gamma_{1\ldots k})\Phi(p_{k},\gamma_{k+1})\\
f_{k,j}(p_{k}\mid\gamma_{1\ldots k+1},\gamma_{k}=0) & =f_{k,j-1}(p_{k}\mid\gamma_{1\ldots k})\\
g_{k,j}(q_{k}\mid\gamma_{1\ldots k+1},\gamma_{k}=0) & \sim g_{k,j-1}(q_{k}\mid\gamma_{1\ldots k})\Phi(q_{k},\gamma_{k+1})\\
g_{k,j}(q_{k}\mid\gamma_{1\ldots k+1},\gamma_{k}=1) & =g_{k,j-1}(q_{k}\mid\gamma_{1\ldots k}).
\end{aligned}
\end{equation}
The $\sim$ indicates normalization, and $f_{0}(p_{k})=1$ and $g_{0}(q_{k})=1$
as before. The distribution $f_{k,j}(p_{k})$ does not change from
$f_{k,j-1}(p_{k})$ unless the space was occupied, and $g_{k,j-1}(q_{k})$
is also left alone unless the space was vacant. In other words, to
update the distributions for a state, a transition out of that state
must have been observed.

\subsubsection{Forgetting Factor}

As training proceeds, the distributions $f_{k}(p_{k})$ and $g_{k}(q_{k})$
become increasingly narrow and converge toward delta functions, the
oldest and newest training data exerting equal but ever-decreasing
influence on the model; even the newest training data becomes diluted.
This is acceptable for batch training, where the history length is
chosen explicitly, but not for incremental training, where eventually
the distributions cannot change at all. We introduce a forgetting
factor $\lambda$ to gradually discount older training data and allow
the Markov chain to retain its flexibility. Linear forgetting is implemented
using
\begin{equation}
\begin{aligned}f_{k,j}^{\prime}(p_{k}) & =\lambda f_{k,j}(p_{k})+(1-\lambda)f_{0}(p_{k})\\
g_{k,j}^{\prime}(q_{k}) & =\lambda g_{k,j}(q_{k})+(1-\lambda)g_{0}(q_{k}),
\end{aligned}
\label{eq:forgetting}
\end{equation}
where $f_{0}(p_{k})=1$ and $g_{0}(q_{k})=1$, and $f_{k,j}(p_{k})$
and $g_{k,j}(q_{k})$ are the posterior distributions that have just
been trained before application of forgetting.%
\footnote{There are other ways to implement forgetting; see \citep{generalForgetting}
for a survey that compares linear with multiplicative forgetting.%
}

There is no direct equivalence between forgetting factors and batch
training history length; batch training (Figure~\ref{fig:trainingSchematic}a)
is analogous to a finite impulse response (FIR) filter with a defined
memory length, while incremental training (Figure~\ref{fig:trainingSchematic}b)
is structurally reminiscent of an infinite impulse response (IIR)
filter where the previous output is fed back into the filter. With
batch training, the hand-picked data set may not contain all the transitions
of interest, so some transitions may not be trained at all. The incremental
approach applies training and forgetting simultaneously, retaining
infrequently observed transitions longer.

\begin{figure}
\begin{centering}
\includegraphics[scale=0.25]{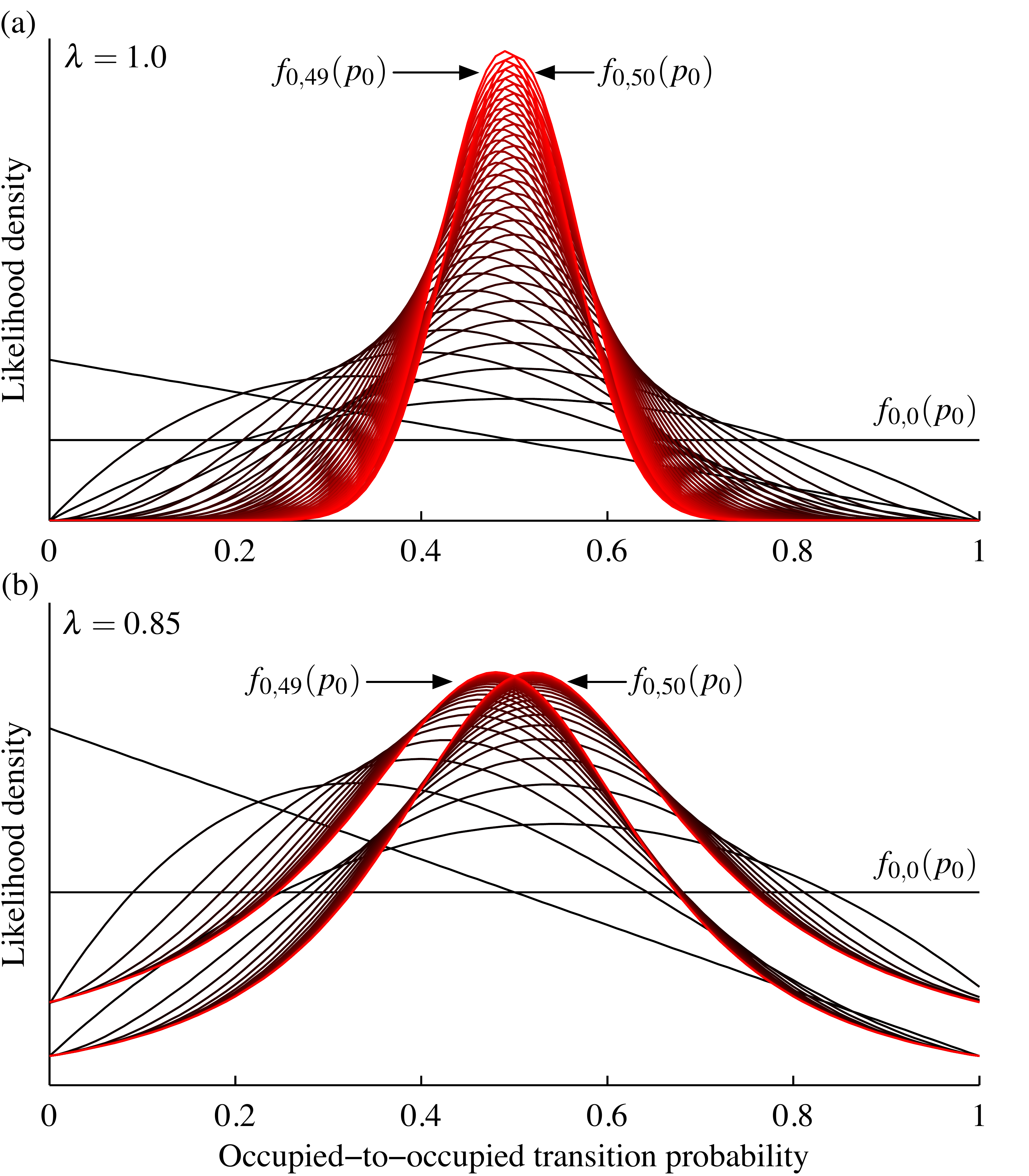}
\par\end{centering}

\protect\caption{\label{fig:forgettingFactor}A particular occupancy transition probability
distribution when trained with alternating data using no forgetting
($\lambda=1$) (a) and with considerable forgetting ($\lambda=0.85$)
(b). The forgetting factor broadens the distribution and allows it
to shift laterally even when extensively trained, reflecting greater
ability to adjust to changes in space usage. For each case, the initial
distribution is uniform (black horizontal trace). Higher levels of
training are shown as brighter color.}
\end{figure}

To illustrate the effect of forgetting on the distributions, we have
trained a single state of the Markov chain repeatedly using alternating
transitions $\gamma_{0}=0\to\gamma_{1}=1$ and $\gamma_{0}=0\to\gamma_{1}=0$.
This is analogous to flipping an unbiased coin numerous times and
observing heads every other flip, from which we expect an increasingly
narrow distribution for $p_{0}$ peaking near 0.5 (Figure~\ref{fig:forgettingFactor}a).
This result would be preferred if the pattern were never expected
to change, but such concentration in the distribution hinders its
ability to change and is therefore undesirable. Using 15\% forgetting
($\lambda=0.85$) gives a distinctly broader distribution lifted off
the horizontal axis (Figure~\ref{fig:forgettingFactor}b). The distribution---and
therefore its expected value---shifts laterally with each alternate
observation, even after many iterations; this extra mobility reflects
greater adaptability. The value $\lambda=0.85$ is much more forgetful
than would be used in practice; Section~\ref{sub:Choosing} will
explore the relationship between $\lambda$ and prediction accuracy.

\subsubsection{Using Fractional Occupancy}

\begin{figure}
\begin{centering}
\includegraphics[scale=0.25]{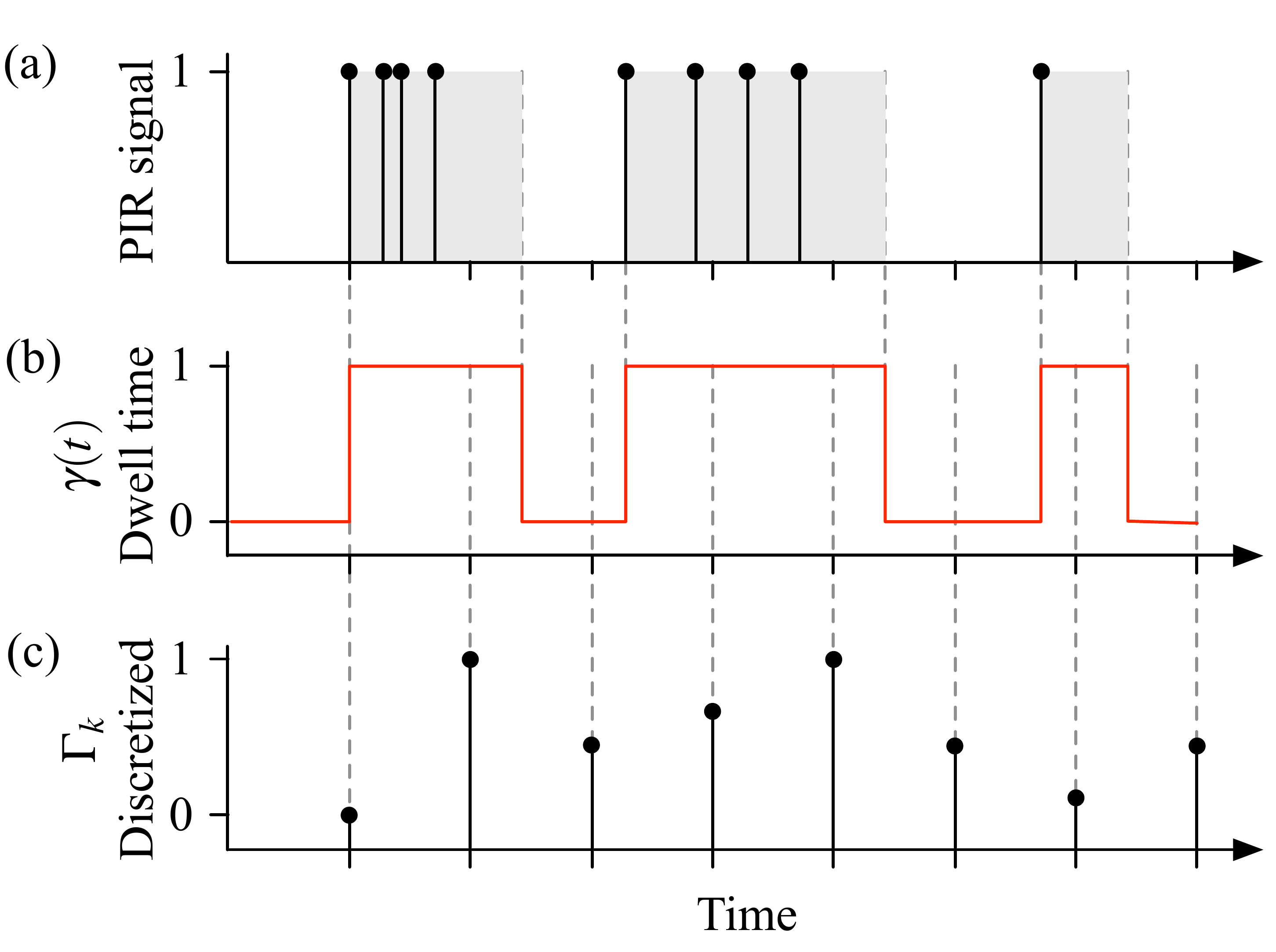}
\par\end{centering}

\protect\caption{\label{fig:discretization}Asynchronous sensor pulses (a); derived
continuous signal using dwell time (b); resulting discrete-time occupancy
percentage (c).}
\end{figure}

Measuring the percentage of occupancy over each time makes occupancy
predictions more precise. To convert the asynchronous pulses from
PIR sensors (Figure~\ref{fig:discretization}a) into a discrete-time
sequence of fractional values, we apply a simple two-step heuristic.
First, we merge closely-spaced pulses using a minimum dwell time to
get a square wave signal $\gamma(t)$ (Figure~\ref{fig:discretization}b).
Then we superimpose a fixed time grid over the signal and average
it over each step to obtain the discrete sequence 
\begin{equation}
\Gamma_{k}=\frac{1}{t_{k}-t_{k-1}}\int_{t_{k-1}}^{t_{k}}\gamma(t)\,\textrm{d}t\equiv\prob{\gamma_{k}=1}\in[0,1],\label{eq:bigGamma}
\end{equation}
essentially treating $\Gamma_{k}$ (Figure~\ref{fig:discretization}c)
as the duty cycle sequence of the pulse width modulated signal $\gamma(t)$.
We subsequently pretend that $\gamma(t)$ is sampled probabilistically
through $\Gamma_{k}$ with a distribution over the Markov state space
$\pi_{k}\in\mathbb{R}^{1\times2M}$. The statements $\Gamma_{k}=60\%$
and $\prob{\gamma_{k}=1}=0.6$ are considered equivalent. 
 From this we estimate the occupancy at time $k+1$ using
\begin{equation}
\begin{alignedat}{2}\prob{\Gamma_{k+1}=1\mid\Gamma_{k}} & = & \Gamma_{k} & \prob{\gamma_{k+1}=1\mid\gamma_{k}=1}\\
 &  & +\,(1-\Gamma_{k}) & \prob{\gamma_{k+1}=1\mid\gamma_{k}=0}\\
 & = & \Gamma_{k} & \ex{p_{k}}+(1-\Gamma_{k})\ex{q_{k}},
\end{alignedat}
\end{equation}
where the expectation operator reflects the fact that $p_{k}$ and
$q_{k}$ are estimated via $f_{k}(p_{k})$ and $g_{k}(q_{k})$. At
each step $k$, there are four possible state transitions with associated
posterior distributions
\begin{equation}
\begin{alignedat}{2}\gamma_{k}=1 & \to\gamma_{k+1}=1:\quad & f_{k,j}^{(1)}(p_{k}) & \sim\Phi(p_{k},1)f_{k,j-1}(p_{k}),\\
\gamma_{k}=1 & \to\gamma_{k+1}=0: & f_{k,j}^{(0)}(p_{k}) & \sim\Phi(p_{k},0)f_{k,j-1}(p_{k}),\\
\gamma_{k}=0 & \to\gamma_{k+1}=1: & g_{k,j}^{(1)}(q_{k}) & \sim\Phi(q_{k},1)g_{k,j-1}(q_{k}),\\
\gamma_{k}=0 & \to\gamma_{k+1}=0: & g_{k,j}^{(0)}(q_{k}) & \sim\Phi(q_{k},0)g_{k,j-1}(q_{k}),
\end{alignedat}
\end{equation}
where $f_{k,j}^{(1)}$ is the updated posterior distribution as if
$\gamma_{k}=1$ and $\gamma_{k+1}=1$ had been observed, $f_{k,j}^{(0)}$
is similar to $f_{k,j}^{(1)}$ but updated as if $\gamma_{k+1}=0$
had been observed, and likewise for $g_{k,j}^{(1)}$ and $g_{k,j}^{(0)}$.
To obtain $f_{k,j}(p_{k})$, we blend $f_{k,j}^{(1)}(p_{k})$ and
$f_{k,j}^{(0)}(p_{k})$ according to the later observation $\Gamma_{k+1}$.
We then weight the training according to $\Gamma_{k}$, which reflects
how likely the space was to have started occupied; values of $\Gamma_{k}$
closer to one apply more training to $f_{k}(p_{k})$, while those
closer to zero cause heavier training of $g_{k}(q_{k})$. The training
for $g_{k,j}(q_{k})$ follows analogously.
\begin{equation}
\begin{alignedat}{2}f_{k,j}(p_{k})\, & =\, & \Gamma_{k}\, & \left(\Gamma_{k+1}f_{k,j}^{(1)}(p_{k})+(1-\Gamma_{k+1})f_{k,j}^{(0)}(p_{k})\right)\\
 & \ + & (1-\Gamma_{k})\, & f_{k,j-1}(p_{k})\\
g_{k,j}(q_{k})\, & =\  & (1-\Gamma_{k})\, & \left(\Gamma_{k+1}g_{k,j}^{(1)}(q_{k})+(1-\Gamma_{k+1})g_{k,j}^{(0)}(q_{k})\right)\\
 & \ + & \Gamma_{k}\, & g_{k,j-1}(q_{k})
\end{alignedat}
\label{eq:fractionalPosteriors}
\end{equation}
Once the new distributions $f_{k,j}(p_{k})$ and $g_{k,j}(q_{k})$
have been found, forgetting is applied similarly to Equation~\ref{eq:forgetting},
where Equations~\ref{eq:fractionalPosteriors} are used instead for
the posterior distributions. The post-forgetting distributions are
then stored.

\subsubsection{Effect of Training on Distribution Shape}

\begin{figure}
\begin{centering}
\includegraphics[scale=0.25]{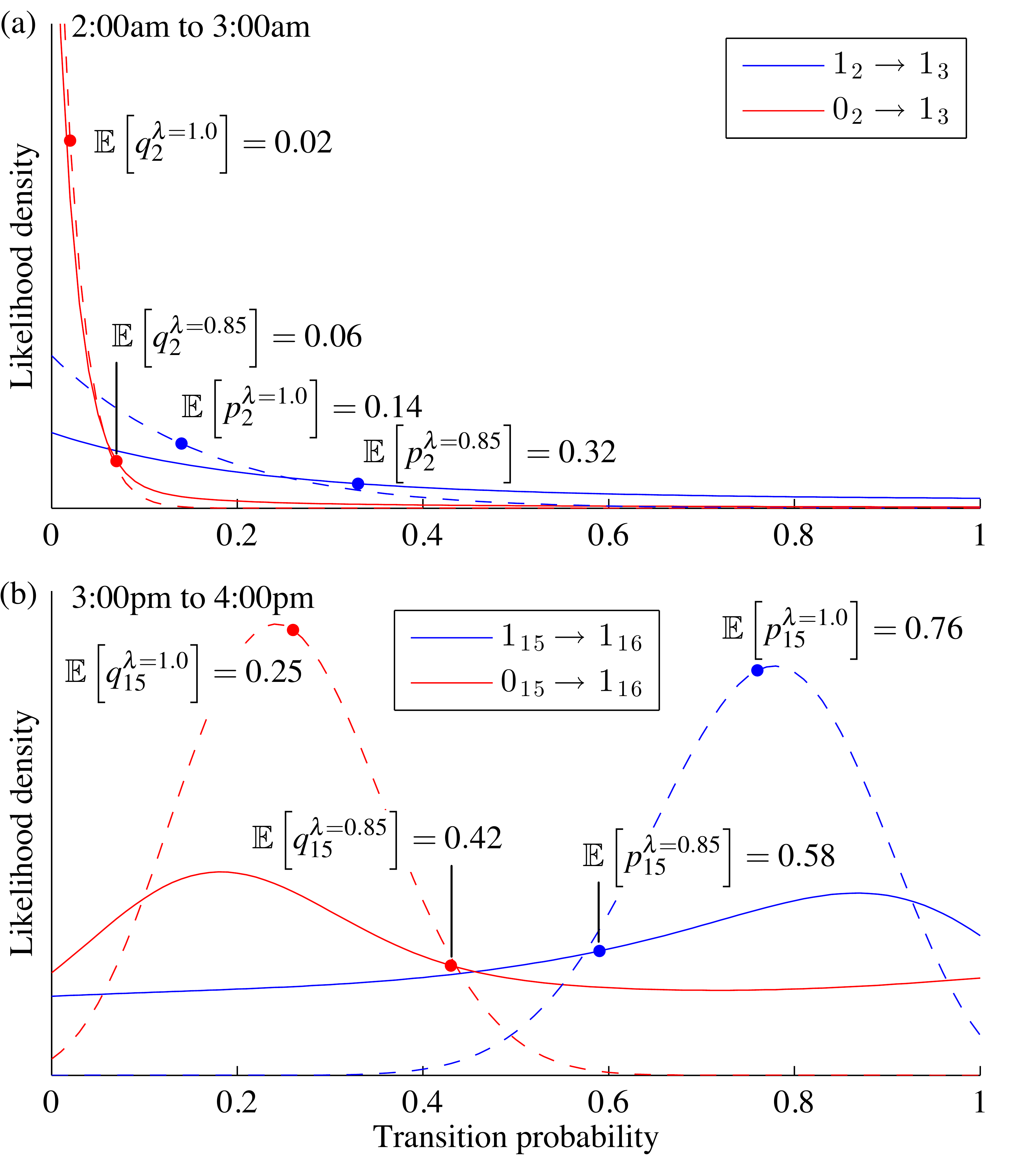}
\par\end{centering}

\protect\caption{\label{fig:Transition-probabilities}Probability densities for the
occupied-to-occupied (blue) and vacant-to-occupied (red) at \formattime{2}{0}{0}
(a) and \formattime{15}{0}{0} (b). Two cases are shown: a heavy forgetting
factor ($\lambda=0.85$, solid) and no forgetting ($\lambda=1.0$,
dashed). }
\end{figure}

To illustrate the connection between training data patterns and the
shapes of $f_{k}(p_{k})$ and $g_{k}(q_{k})$, we have trained two
Markov chains with the MERL Belady conference room data from March
22 to June 9 and sampled the distributions afterward. In Figure~\ref{fig:Transition-probabilities},
two sets of distributions---one for \formattime{2}{0}{0}$\to$\formattime{3}{0}{0}
(a) and the other for \formattime{15}{0}{0}$\to$\formattime{16}{0}{0}
(b)---are shown for both strong forgetting ($\lambda=0.85$, solid)
and no forgetting ($\lambda=1.0$, dashed). In Figure~\ref{fig:Transition-probabilities}a
we see that both occupancy and vacancy at \formattime{2}{0}{0} strongly
imply vacancy at \formattime{3}{0}{0}. In other words, early morning
occupancy is very uncommon and usually brief. Because occupancy is
rare at \formattime{2}{0}{0}, the transition $\gamma_{2}=1_{2}\to\gamma_{3}=1_{3}$
(blue) is very weakly trained and has a very flat distribution. In
Figure~\ref{fig:Transition-probabilities}b, we see that occupancy
at \formattime{15}{0}{0} is more varied, resulting in more typical
bell-shaped distributions. The distributions for \formattime{15}{0}{0}
suggest that meetings are likely to continue into the next hour but
are unlikely to start the following hour. The distributions for $\lambda=0.85$
are shaped similarly to those for $\lambda=1.0$ but are markedly
subdued with expected values closer to $0.5$.

\subsection{Transition Matrix and Occupancy Prediction}

Recall from Section~\ref{sub:Markov-chain-formulation} and Figure~\ref{fig:markovChain}b
the Markov chain has states $1_{0}\ldots1_{23}$ and $0_{0}\ldots0_{23}$.
The probability distribution of the current occupancy state $\pi_{k}\in\mathbb{R}^{1\times2M}$
evolves according to $\pi_{k+1}=\pi_{k}P$. The matrix $P$ can be
constructed from the four blocks: $P^{(\textrm{\Romannum{1})}}$ for
$1_{k}\to1_{k+1}$, $P^{(\textrm{\Romannum{2}})}$ for $1_{k}\to0_{k+1}$,
$P^{(\textrm{\Romannum{3}})}$ for $0_{k}\to1_{k+1}$, and $P^{(\textrm{\Romannum{4}})}$
for $0_{k}\to0_{k+1}$ transitions. The entries for $P^{(\textrm{\Romannum{1}})}$
and $P^{(\textrm{\Romannum{4}})}$ are the expected values of $p_{0\ldots M-1}$
and $q_{0\ldots M-1}$, and the other two matrices are their complements.
\newlength{\foo}
\setlength{\foo}{4.5cm}
\begin{equation}
\begin{aligned}
P_{ik}^{\textrm{(\Romannum{1})}} &=
  \begin{cases}
    \makebox[\foo][l]{$\prob{\gamma_k=1\mid\gamma_i=1}=\ex{p_i}$} & (\dagger) \\
    0 & \textrm{otherwise}
  \end{cases}\\
P_{ik}^{\textrm{(\Romannum{2})}} &= 
\begin{cases}
    \makebox[\foo][l]{$\prob{\gamma_k=0\mid\gamma_i=1}=1-\ex{p_i}$} & (\dagger) \\
    0 & \textrm{otherwise}
  \end{cases}\\
P_{ik}^{\textrm{(\Romannum{3})}} &= 
\begin{cases}
    \makebox[\foo][l]{$\prob{\gamma_k=1\mid\gamma_i=0}=\ex{q_i}$} & (\dagger) \\
    0 & \textrm{otherwise}
  \end{cases}\\
P_{ik}^{\textrm{(\Romannum{4})}} &= 
\begin{cases}
    \makebox[\foo][l]{$\prob{\gamma_k=0\mid\gamma_i=0}=1-\ex{q_i}$} & (\dagger) \\
    0 & \textrm{otherwise}
  \end{cases}
\end{aligned}
\end{equation}where $\dagger$ means $k=i+1\ (\bmod M)$. For example, $P^{\textrm{(\Romannum{1})}}$
takes the form
\begin{equation}
P^{(\textrm{\Romannum{1}})}=\exnobr{\begin{bmatrix}0 & p_{0}\\
 & 0 & p_{1}\\
 &  & \ddots & \ddots\\
 &  &  & 0 & p_{M-2}\\
p_{M-1} &  &  &  & 0
\end{bmatrix}}.
\end{equation}
The complete matrix is
\begin{equation}
P=\left[\begin{array}{c|c}
P^{\textrm{(\Romannum{1})}} & P^{\textrm{(\Romannum{2})}}\\
\hline \vertTenPt P^{\textrm{(\Romannum{3})}} & P^{\textrm{(\Romannum{4})}}
\end{array}\right].
\end{equation}
The expected occupancy $m$ steps in the future given a current estimate
$\Gamma_{k}$ is
\begin{equation}
\ex{\Gamma_{k+j}\mid\Gamma_{k}}=\underbrace{\begin{bmatrix}\Gamma_{k}\one_{1\times M}^{k} & (1-\Gamma_{k})\one_{1\times M}^{k}\end{bmatrix}}_{\pi_{k}}P^{j}\begin{bmatrix}1_{M\times1}\\
0_{M\times1}
\end{bmatrix}\label{eq:occupancyProjection}
\end{equation}
where $\one_{1\times M}^{k}$ is a vector with the $k$th element
set to one and all others left zero.

\section{MPC Formulation}

To balance competing demands for occupant comfort and low total energy
consumption, we need to avoid conditioning the space when vacancy
is expected; the level of comfort should scale with occupancy. To
simplify the cost function, we have augmented the building's state
space model with the non-changing temperature setpoint and a weather
forecast shift-register system, i.e.
\begin{equation}
\underbrace{\begin{bmatrix}x_{k+1}\\
\tau_{k+1}\\
\phi_{k+1}
\end{bmatrix}}_{\tilde{x}_{k+1}}=\tilde{A}\underbrace{\begin{bmatrix}x_{k}\\
\tau_{k}\\
\phi_{k}
\end{bmatrix}}_{\tilde{x}_{k}}+\underbrace{\begin{bmatrix}B_{u}\\
0\\
0
\end{bmatrix}}_{\tilde{B}}u_{k},
\end{equation}
where $x$ is the building's thermal state (Equation~\ref{eq:buildingStateEquation}),
$\tau$ is the comfort setpoint, and $\phi$ is a shift-register state
that iterates through the weather forecast over the MPC horizon. The
augmented matrix $\tilde{A}$ connects the weather forecast to the
building thermal model internally.%
\footnote{Because the forecast is updated at each time step, our implementation
adjusts the augmented transition matrix $\tilde{A}$ before each MPC
synthesis to reflect the latest prediction. This simplifies the cost
function and allows the MPC to be formulated in a compact vectorized
form as detailed in Equation~3.8 of \citep{rossiter2003model}.%
} We seek the optimal control law
\begin{equation}
\begin{alignedat}{2}u_{k}^{*}(\tilde{x}_{k},\Gamma_{k}) & = & \underset{u}{\arg\min}\quad & \ex{\sum_{j=0}^{N-1}g(\tilde{x}_{k+j},u_{k+j},\Gamma_{k+j})}\\
 &  & \textrm{subject to}\quad & \begin{aligned}\tilde{x}_{i+1} & =\tilde{A}\tilde{x}_{i}+\tilde{B}u_{i}\quad\forall i\in\mathbb{Z}^{+}\\
0 & \leq u\leq u_{\max}
\end{aligned}
\end{alignedat}
\end{equation}
where $u_{k+j}$ is an individual control action and $\Gamma_{k+j}\in[0,1]$
is an occupancy measurement or prediction. This is standard except
that the stage cost adjusts the discomfort weigh using occupancy,
i.e. 
\begin{equation}
\begin{aligned}g(\tilde{x},u,\Gamma) & =\tilde{x}^{\top}\Gamma Q\tilde{x}+r\left|u\right|\\
 & =\Gamma\beta\left(x_{\textrm{zone}}-\tau\right)^{2}+r\left|u\right|,
\end{aligned}
\end{equation}
where
\begin{itemize}
\item $\tilde{x}$ is the augmented system state vector and $x_{\textrm{zone}}$
is the zone air temperature being controlled,
\item $u$ is the heat input to the zone,
\item $\Gamma$ is the observed or predicted occupancy,
\item $\tau$ is the comfortable setpoint temperature (constant),
\item $Q$ is a matrix that extracts $\beta(x_{\textrm{zone}}-\tau)^{2}$
from $\mbox{\ensuremath{\tilde{x}}}^{\top}Q\tilde{x}$, and
\item $\beta$ and $r$ are the discomfort and energy cost gains.%
\footnote{In this simplified formulation, only the ratio between $r$ and $\beta$
matters; together, they constitute a single tuning adjustment. The
energy cost gain $r$ can be time-varying if one wishes, for example,
to incorporate time-of-day utility pricing.%
}
\end{itemize}
The many ($2^{N}$) possible occupancy state trajectories, along
with the constraints on $u$, make it difficult to find a closed-form
solution using exact dynamic programming. (Recall from Figure~\ref{fig:markovChain}
that each occupancy state has two possible outgoing transitions.)
If we instead condition all occupancy predictions solely on the present
observation, we obtain the approximation
\begin{equation}
\begin{aligned}u_{k}^{*}(\tilde{x}_{k},\Gamma_{k}) & \approx\underset{0\leq u\leq u_{\max}}{\arg\min}\Biggl\{ g(\tilde{x}_{k},u_{k},\Gamma_{k})\\
 & +\sum_{j=1}^{N-1}g\left(\tilde{x}_{k+j},u_{k+j},\mathbb{E}\left[\Gamma_{k+j}\mid\Gamma_{k}\right]\right)\Biggr\},
\end{aligned}
\end{equation}
 where $\ex{\Gamma_{k+j}\mid\Gamma_{k}}$ comes from Equation~\ref{eq:occupancyProjection}.%
\footnote{Multi-parametric methods can be used to partition the state space
into regions, each with an exact control law parameterized on the
entire state at the expense of a more complex MPC formulation \citep{multiParametricProgramming}.%
} The optimization is then
\begin{equation}
\begin{aligned}\min_{u_{k}\cdots u_{k+N-1}}\quad & \sum_{j=0}^{N-1}\tilde{x}_{k+j}^{\top}\ex{\Gamma_{k+j}\mid\Gamma_{k}}Q\tilde{x}_{k+j}+r\left|u_{k+j}\right|\\
\textrm{subject to\quad} & \begin{aligned}\tilde{x}_{i+1} & =\tilde{A}\tilde{x}_{i}+\tilde{B}u_{i}\quad\forall i\in\mathbb{Z}^{+}\\
0 & \le u\le u_{\max}
\end{aligned}
\end{aligned}
\end{equation}
As with conventional MPC, the controller applies $u_{k}$ to the system
and discards $u_{k+1}\ldots u_{k+N-1}$; the solution is repeated
at each subsequent step. 

\begin{figure}
\begin{centering}
\includegraphics[scale=0.25]{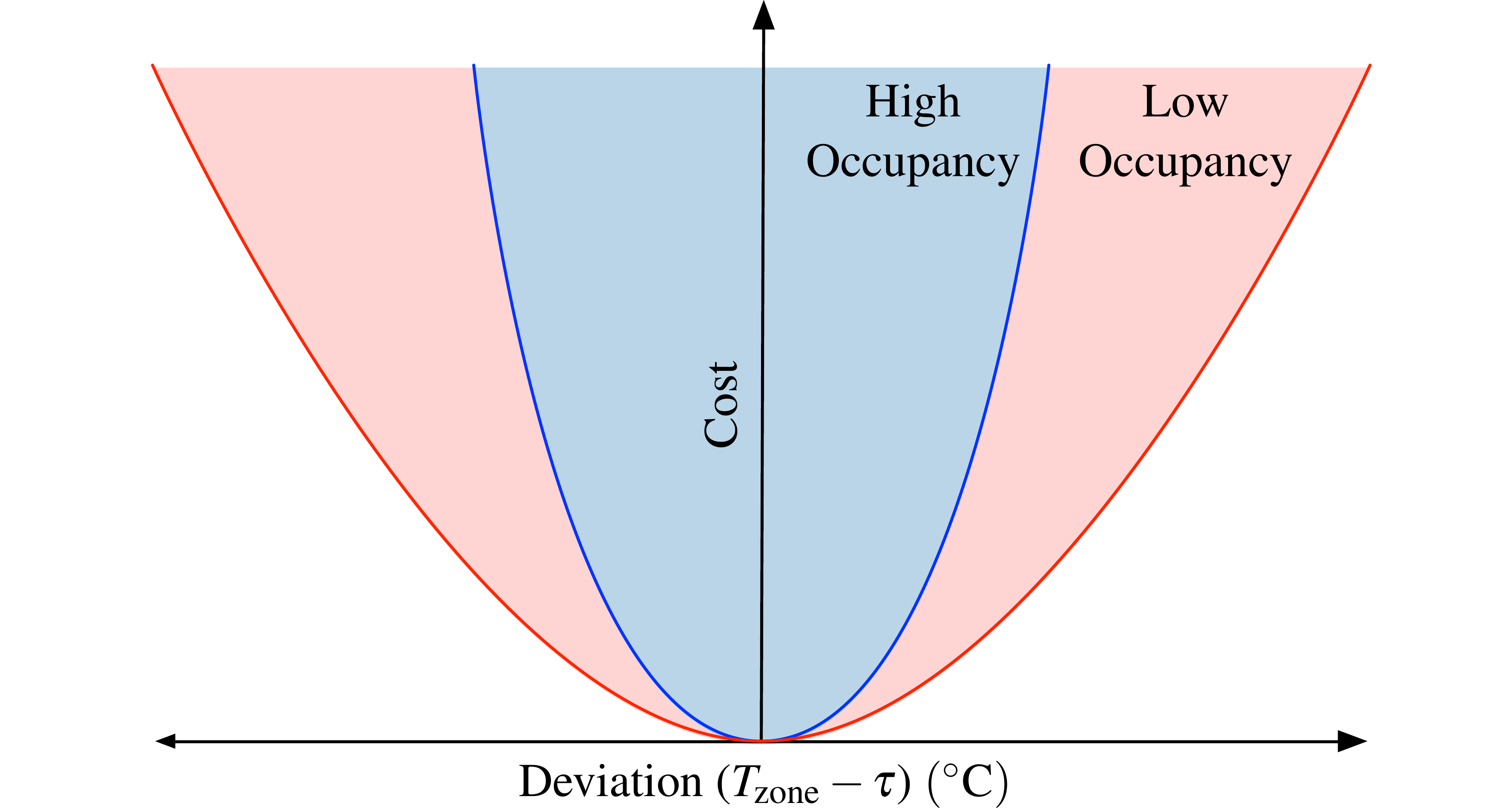}
\par\end{centering}

\protect\caption{\label{fig:Cost-function}Discomfort cost for high expected occupancy
(blue) and low expected occupancy (red). When high occupancy is predicted,
the curve steepens and less deviation from the setpoint is permitted.}
\end{figure}

The controller never reaches the setpoint $\tau$ for two reasons.
First, including energy in the cost function counteracts temperature
regulation, with the trade-off tuned through the ratio $\beta/r$.
Second, the discomfort cost is weighted by \emph{expected} occupancy,
which never reaches 1.0. We have chosen to penalize $\left|u\right|$,
rather than $u^{2}$, because quadratic cost suppresses peaks and
spreads control action over time; peak suppression inhibits the full
system shutdown necessary to save energy during vacancy. When high
occupancy is predicted, the discomfort cost (Figure~\ref{fig:Cost-function})
becomes steeper and causes the temperature to more closely approach
the setpoint. 

\section{Comparison to Conventional Control}

\subsection{Experimental Setup}

To demonstrate the algorithm's advantages over conventional control,
we have run a simulation under the following conditions:
\begin{itemize}
\item MERL occupancy data for the Belady conference room (sensors 452 and
453) from February 12 to April 10, 2007;
\item EnergyPlus weather data for Elmira, NY starting March 1 (typical meteorological
year) and a three-week warm-up period;
\item no un-modeled disturbances;
\item one-hour time step;
\item system capacity of 8.0kW.
\end{itemize}
The thermal model is the single-zone building RC network discussed
previously. To emphasize the benefit of prediction, we have chosen
the weather period to just saturate the control output in typical
winter conditions. (These conditions emphasize the need to predict
more than one hour out; we could have chosen January and increased
the system capacity slightly for the same result.)

\subsection{Choosing $\lambda$\label{sub:Choosing}}

\begin{figure}
\begin{centering}
\includegraphics[scale=0.25]{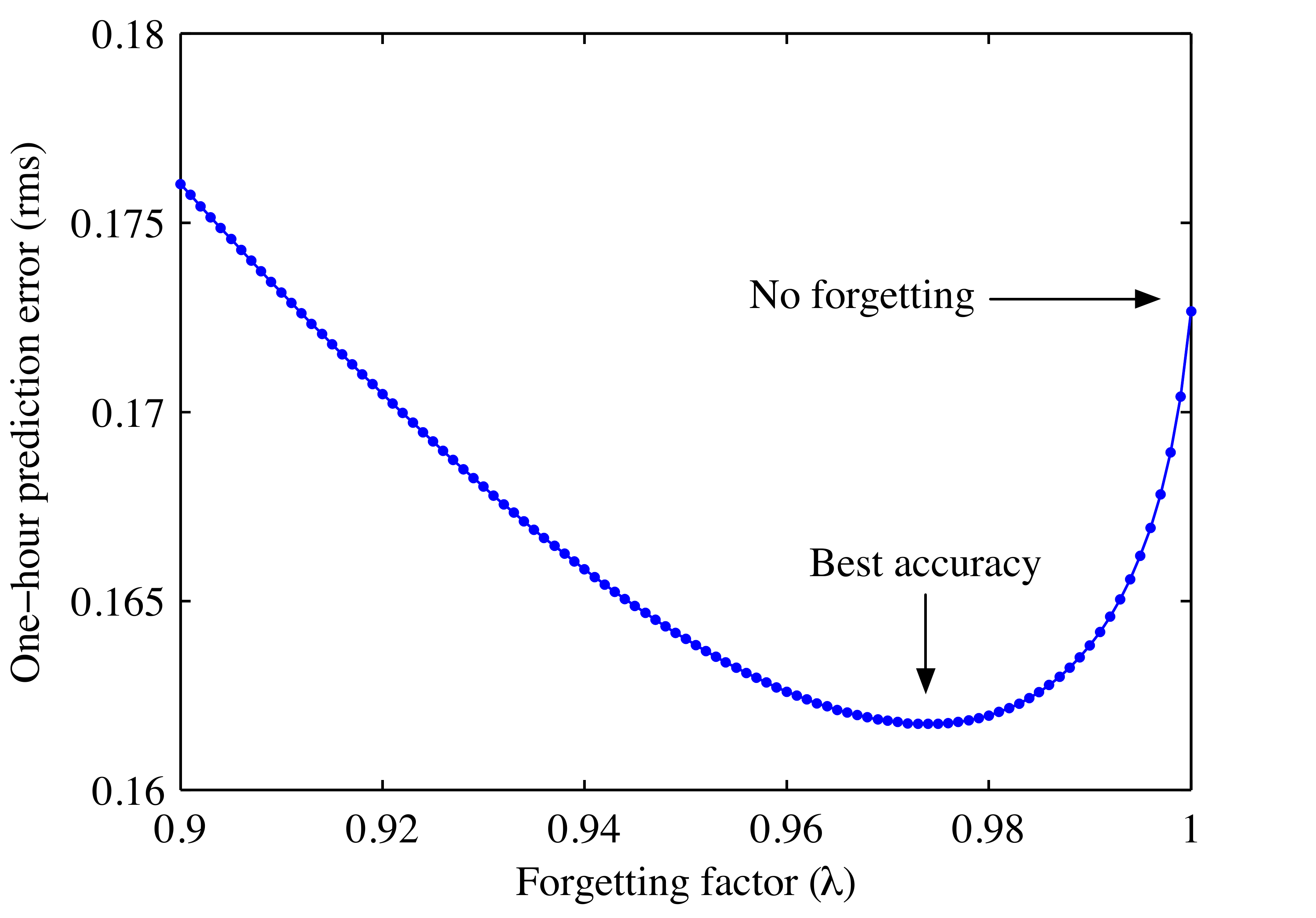}\protect\caption{\label{fig:Influence-of-forgetting}Influence of forgetting factor
$\lambda$ on one-hour root-mean-square prediction error. These results
were obtained by training the model incrementally over the MERL Belady
conference room occupancy data from February 12 to April 10, 2007
and simultaneously comparing each observation to the prediction made
in the previous hour.}

\par\end{centering}

\end{figure}

Before we run the simulation, we need to choose the forgetting factor.
Without forgetting ($\lambda=1.0$), consistent occupancy patterns
allow predictions to asymptotically approach $\Gamma=0$ and $\Gamma=1$,
but the ever-lengthening effective history length hinders adaptation
and leads to very large prediction error. At the other extreme, high
forgetting ($\lambda\ll1.0$) gives a model easily distracted by irregularities
that consistently predicts occupancy near $\Gamma=0.5$, which again
leads to high prediction error. Intuition suggests that a minimum
prediction error should exist between these limits, and indeed this
is the case. Figure~\ref{fig:Influence-of-forgetting} shows the
relationship between $\lambda$ and one-hour prediction error using
the simulation occupancy data, with $\lambda=0.974$ giving the best
prediction accuracy. Of course, there is no guarantee that the best
past value of $\lambda$ will work well in the future; nonetheless,
the convexity suggests that $\lambda$ could be calibrated on-line
with an extremum-seeking algorithm \citep{ariyur2003real}. 

\subsection{Performance Comparison}

Figure~\ref{fig:Simulation-results} shows simulation results for
three identically-tuned MPC implementations:
\begin{enumerate}
\item a purely occupancy-triggered controller,
\item a scheduled controller supplemented with occupancy triggering, and
\item an on-line trained occupancy-predicting controller with one week of
pre-training ($\lambda=97.4\%$).
\end{enumerate}
The occupancy-triggered controller (green) maintains $\tau=23^{\circ}\textrm{C}$
during occupied hours and $10^{\circ}\textrm{C}$ during vacant hours.
The scheduled controller uses the same setpoint from \formattime{5}{0}{0}
to \formattime{21}{0}{0} and any time the space is occupied. To simplify
the simulation, all three controllers ignore occupancy and control
to $10^{\circ}\textrm{\textrm{C}}$ ($50^{\circ}\textrm{F}$) over
weekends.

\subsubsection{Energy and Comfort}

\begin{figure*}
\begin{centering}
\includegraphics[scale=0.2]{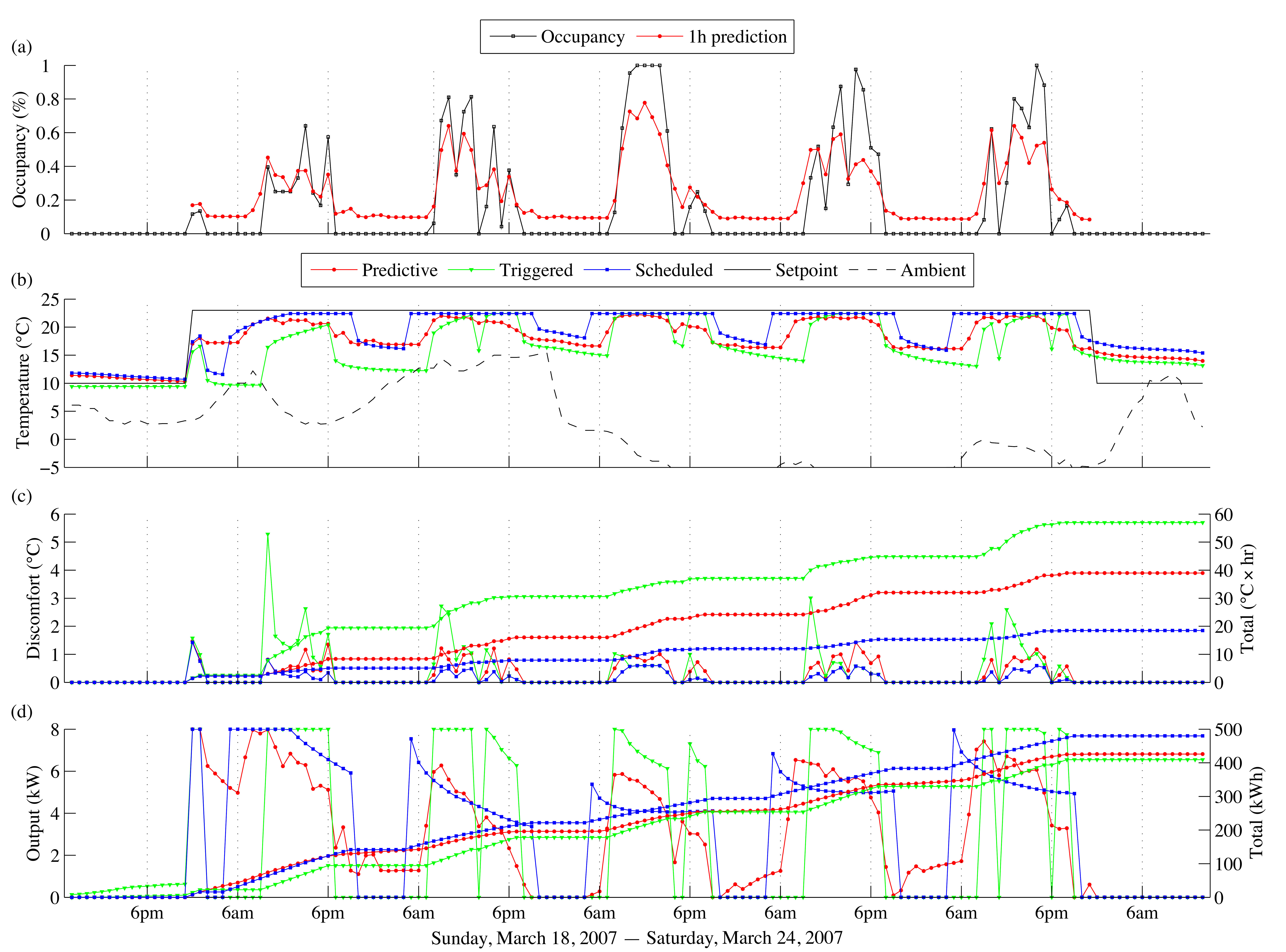}
\par\end{centering}

\protect\caption{\label{fig:Simulation-results}Simulation results for a single-zone
building: occupancy prediction (a), temperature control performance
and ambient conditions (b), occupant discomfort (c), and energy consumption
(d). The discomfort (deviation from setpoint) is weighted by the occupancy
$\Gamma_{k}$.}
\end{figure*}

The occupancy-triggered controller consumes by far the least energy
because it does not account for thermal lag or expected occupancy
and therefore runs the least. Not surprisingly, its comfort performance
upon occupant arrival is very poor, with large leading spikes on the
discomfort trace in Figure~\ref{fig:Simulation-results}c and frequent
calls for maximum output power in Figure~\ref{fig:Simulation-results}d.
The scheduled controller leaves plenty of margin around the typical
occupancy envelope and consequently yields excellent comfort at the
expense of energy efficiency. The comfort performance of occupancy
predicting MPC lies between these two methods, with peak discomfort
slightly worse than scheduled control but without the severe deviations
of triggered control. Table~\ref{fig:resultTable} shows up to 19\%
energy savings compared to the scheduled controller and significantly
lower peak discomfort than the occupancy-triggered controller.

\begin{figure}
\begin{centering}
\includegraphics[scale=0.25]{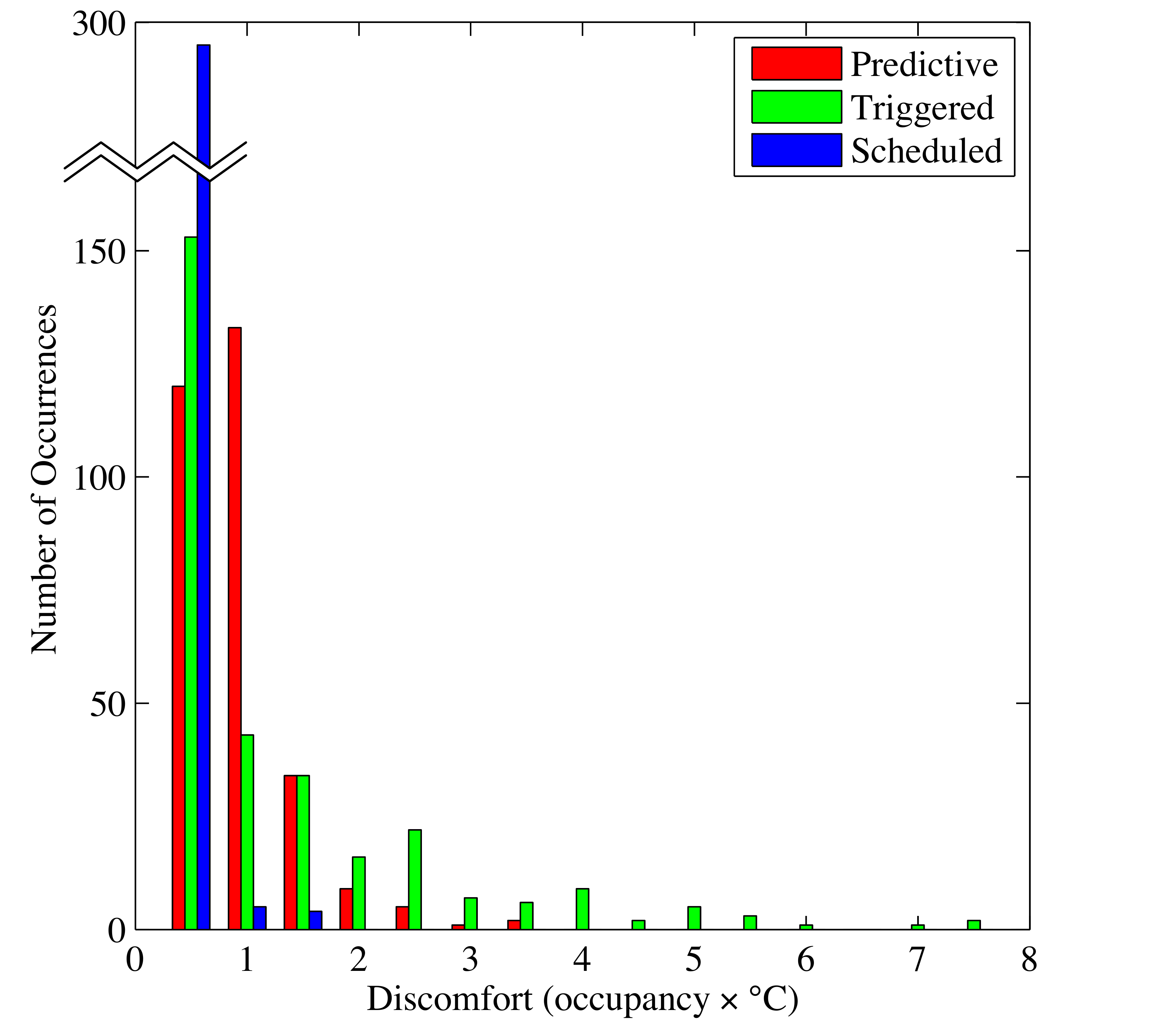}
\par\end{centering}

\protect\caption{\label{fig:histogram}Distribution of occupancy-weighted discomfort
over a two-month simulation. The properly-tuned schedule shows very
little discomfort over the two-month simulation, while occupancy-triggered
control produces many severe instances of discomfort. Occupancy predicting
control yields a distribution similar to that of scheduled control
but shifted slightly to the right.}
\end{figure}

Perhaps more interesting than the discomfort peak is its distribution.
Figure~\ref{fig:histogram} shows how many times various occupancy-weighted
discomfort levels occur under each control method. It comes as little
surprise that the scheduled controller maintains discomfort within
$2^{\circ}\textrm{C}$ at all times. (Clearly, though, an out-of-date
schedule would not perform this well, so this is a rather optimistic
profile of scheduled control.) The occupancy-predicting controller
maintains discomfort less than $2^{\circ}\textrm{C}$ more than 94\%
of the time with relatively mild outliers. The occupancy-triggered
controller trails with 75\% incidence of low discomfort and numerous
severe violations. In summary, the occupancy predicting control scheme
yields comfort performance that rivals that of a properly tuned schedule.

Energy performance is also as expected. The conservative schedule
leaves ample time to pre-condition the space along with some margin
in the evening. The cost of this performance is 24\% more total energy
over the simulation than the occupancy-triggered controller. Consumption
by the occupancy-predicting controller is moderate, at 12\% more than
the triggered and 19\% less than the scheduled control, and there
are very few instances where the system needs to run at maximum power
to catch up.

\begin{table}
\begin{centering}
\footnotesize{}%
\begin{tabular}{cccccc}
\hline 
 & \multicolumn{3}{c}{Discomfort ($^{\circ}\textrm{C\ensuremath{\times\textrm{hr\ensuremath{\times\textrm{occ.}}}}}$)} & \multicolumn{2}{c}{Energy}\tabularnewline
 & Total & Peak & Variance & Total (kWh) & Savings (\%)\tabularnewline
\hline 
Predictive & 270 & 3.73 & 0.20 & 2493 & 19\tabularnewline
Triggered & 396 & 7.67 & 0.72 & 2237 & 27\tabularnewline
Scheduled & 108 & 1.69 & 0.04 & 3088 & 0\tabularnewline
\end{tabular}
\par\end{centering}

\protect\caption{\label{fig:resultTable}Predictive, triggered, and scheduled control
performance summary for two-month simulation. }
\end{table}

\section{Conclusion}

We have demonstrated the use of model predictive control with a stochastic
occupancy model to reduce HVAC energy consumption. Using occupancy
predicted by an automatically-trained Markov chain, the algorithm
is simplified by approximate dynamic programming where occupancy is
projected multiple steps into the future using a current observation.
We remark that although our method relies on weather forecasts and
a dynamical model of the building, on-line data sources and emerging
software tools have made these easy to acquire. 

We have made some simplifications to improve clarity. First, we have
chosen a rather coarse one-hour time step, even though practical controllers
normally operate on a much finer time scale to provide adequate bandwidth;
the Markov model may, however, operate on an entirely different time
scale from the MPC with only minor implementation changes. Second,
our hypothetical system has constant efficiency and operates only
in heat mode to simplify the cost function and maintain focus on the
paper's contribution. As long as energy consumption can be controlled
and room temperature can be measured, the stochastic occupancy model
may be applied to arbitrarily complex {MPC} scenarios. Finally,
we have used a certainty-equivalence assumption for weather and occupancy
predictions; recent research has addressed ways to incorporate uncertainty
into the optimization for added robustness. Demonstrating our algorithm
without these simplifications is left to future work.

\section{Acknowledgements}

The authors thank Peter Radecki for his constructive feedback.

\section{References}

\bibliographystyle{plain}
\bibliography{library,mypubs}

\end{document}